# Game Theory based Joint Task Offloading and Resources Allocation Algorithm for Mobile Edge Computing

Jianen Yan, Ning Li, *Member, IEEE*, Zhaoxin Zhang, Alex X. Liu, *Fellow, IEEE*, Jose Fernan Martinez, Xin Yuan

*Abstract*-Mobile edge computing (MEC) has emerged for reducing energy consumption and latency by allowing mobile users to offload computationally intensive tasks to the MEC server. Due to the spectrum reuse in small cell network, the inter-cell interference has a great effect on MEC's performances. In this paper, for reducing the energy consumption and latency of MEC, we propose a game theory based approach to join task offloading decision and resources allocation together in the MEC system. In this algorithm, the offloading decision, the CPU capacity adjustment, the transmission power control, and the network interference management of mobile users are regarded as a game. In this game, based on the best response strategy, each mobile user makes their own utility maximum rather than the utility of the whole system. We prove that this game is an exact potential game and the Nash equilibrium (NE) of this game exists. For reaching the NE, the best response approach is applied. We calculate the best response of these three variables. Moreover, we investigate the properties of this algorithm, including the convergence, the computational complexity, and the Price of anarchy (PoA). The theoretical analysis shows that the inter-cell interference affects on the performances of MEC greatly. The NE of this game is Pareto efficiency. Finally, we evaluate the performances of this algorithm by simulation. The simulation results illustrate that this algorithm is effective in improving the performances of the multi-user MEC system.

*Keywords*-Mobile edge computing, task offloading, transmission power control, game theory, interference.

## I. INTRODUCTION

### A. Motivation and Problem Statement

The task offloading and the resources allocation are important to the performances of multi-user MEC [1-4]. First, an appropriate offloading decision and resources allocation strategy can improve the performances of multi-user MEC greatly, such as reducing energy consumption and execution latency. Second, since the network resources, such as energy, memory space, computing capacity, is limited in both MEC server and mobile devices, it is efficient to choose an appropriate task offloading decision and resource allocation strategy to make the system utility maximum. The task offloading means that the mobile users choose to offload computation task to MEC server or compute locally based on the energy consumption and execution latency of this task. The resources allocation means that the MEC server or the mobile user chooses appropriate network resources (such as transmission power, CPU capacity, communication channel, etc.) to reduce the energy consumption and execution latency of MEC [8-15]. Moreover, since the offloading decision and the resources allocation are related and can affect each other, the most effective approach is to optimize these two items jointly rather than separately. There are two different kinds of approaches to achieve the above goals: the traditional optimization approach (such as convex optimization [8-15]) and the game theory based approach [5-6]. Until now, the traditional optimization has been well studied. However, there are some limitations in this approach, such as the heavy burden of complex information exchange between MEC server and mobile users, high computational complexity, etc. Even the game theory based approach can overcome the disadvantages of the traditional optimization approach (which will be explained in detail in the rest of this paper), the research on game theory based approach is just starting. There are still many issues need to be investigated, such as power control, interference management, etc.

Therefore, the fundamental problem will be investigated in this paper can be summarized as: how can we join the offloading decision and the resource allocation together based on game theory to improve the performances of MEC, i.e., reducing energy consumption and execution latency? This is challenging since the offloading decision and the resources allocation are related and can affect each other [24].

### B. Limitations of Prior Arts

There are some limitations on both the traditional optimization approach and the game theory based approach [5-6]. *First*, even the traditional optimization approach is effective in improving the performances of MEC, it has three nonnegligible disadvantages [5-6]: (1) because the optimization objective in traditional optimization approach is the utility of the whole system, so there have massive information collection from mobile devices and massive information exchange between mobile devices and MEC server; (2) since different mobile devices are usually owned by different individuals and they may pursue different interests, the traditional optimization approach has limited capability to reflect the interaction between mobile users; (3) the computational complexity of traditional optimization approach is high; on the one hand, the optimization objective is complex; on the other hand, since the optimization objective is complex, the calculation of optimal solutions is also complicated, such as decoupling, convex optimization, etc. *Second*, game theory based approaches are proposed in recent years. However, on the one hand, the research on game theory based approach is just starting; such as the algorithms proposed in [5] and [6]; on the other hand, at present, the game theory based approach only takes offloading decision and channel allocation into account. This is far from enough for the MEC system.

### C. Proposed Approach and Advantages over Prior Arts

In this paper, we propose a game theory based approach to join task offloading decision and resources allocation together in the MEC system. In this algorithm, the offloading decision, the CPU capacity adjustment, the transmission power control, and the network interference management of mobile users are regarded as a game. In this game, based on the best response strategy, each mobile user makes their own utility maximum rather than the utility of the whole system. Based on this algorithm, the system can reach the state of equilibrium and the performance can be optimized.

The proposed algorithm addresses the limitations of prior arts, which can be summarized as follows. *Compared with the traditional optimization approaches*, the proposed solution has the same capability on dealing with multi-parameters, i.e., joint optimizing offloading decision, transmission power, and CPU capability under interference environment in multi-user MEC system; however, our proposed algorithm has low computational complexity and information exchange, so the disadvantages of traditional optimization approach can be overcame. *Compared with the game theory based algorithms* (i.e., the algorithms that proposed in [5] and [6]), except offloading decision, channel allocation, and CPU capability, our proposed algorithm can also take the transmission power control into account in interference-aware multi-user MEC system; thus, our proposed algorithm has much better performances than that presented in [5] and [6].

*D. Technical Challenges and Solutions*

There are some technical challenges to solve the problems presented in Section I.A. *The first technical challenge* is to join offloading decision, CPU capacity adjustment, transmission power control, and channel allocation together in a game. This is a technical challenge because of two reasons: (1) the feasible regions of communication channel and transmission power are not continuous; the communication channel is not continuous is obvious; however, due to the interference, the transmission power should make the Signal to Interference plus Noise Ratio (SINR) at MEC server larger than threshold; thus, the feasible region of transmission power should be $p \in \{0, [p_{min}, p_{max}]\}$; (2) these parameters are related, so it is complicated for the game to reach NE. To address this challenge, first, we introduce potential game into our algorithm; based on the theory proposed in [19-20, 28], we take the neighbors' utilities into account to construct potential function and prove that the NE of this game exists; second, we introduce the best response strategy into our algorithm to reach NE. During the execution process, first of all, the mobile users get the channel interference information from the Base station (BS); based on the channel interference information, each mobile user calculates the best response of offloading decision, transmission power, and CPU capacity locally; each mobile user adjusts their offloading decision, transmission power, and CPU capacity to these values. This process will be repeated until the NE of this game is reached. *The second technical challenge* is to calculate the best response of offloading decision, transmission power, and CPU capacity. This is a technical challenge because of the interaction between different users, which makes the calculation difficult. To address this technical challenge, in this paper, we propose a set of algorithms to calculate the best response of these four parameters. *The third technical challenge* is to prove the proposed algorithm is effective and can reach NE. This is a technical challenge since the proposed game theory based algorithm takes more parameters into account than the previous works, which makes the analysis of the proposed algorithm difficult. To address this challenge, in this paper, we have proved that: (1) our proposed game theory based algorithm is convergent, and the NE of this game exists and is Pareto efficiency; (2) our proposed algorithm is Polynomial Local Search complete, which means that our algorithm can be finished in polynomial time; (3) the computational complexity of our proposed algorithm is $O[Cn \log^3(n)]$; (4) reducing the interference between mobile users can improve the performances of MEC effectively.

## II. RELATED WORKS

The main purpose of MEC is to reduce energy consumption and latency by carefully designing the task offloading scheme and resources allocation manner [1-4], which has been learned separately or jointly in both the single-user and multi-user MEC system [6-18]. In [6], the author proposes a game theory approach which can achieve efficient computation offloading for mobile cloud computing; the authors in [5] apply the conclusions in [6] into the multi-user MEC system and propose a game theory based offloading decision algorithm; in this algorithm, the mobile users decide whether to offload the tasks to MEC servers and which communication channels are used in a distributed manner. However, in these two algorithms, only the interference, the CPU capability, and the offloading decision are considered. Even the authors in these two papers propose that power control is a potential approach to minimize network interference and improve the performance of MEC, they fail to find a solution to overcome this problem. In [7], the authors propose an optimal offloading policy for the applications which include sequential component dependency graphs and multi-radio enabled mobile devices. Different from the algorithm shown in [7], the authors in [8] solve the task offloading issues for arbitrary dependency graphics. In [9], the authors propose an effective computation model for the MEC system, which joints the computation and communication cooperation together to improve the system performances.

As the points of view that are proposed in [1] and [2], not only the offloading decision but also resources management is important for improving the performance of the MEC system. Resources management includes communication channel allocation, CPU capability control, transmission power control, etc. [10]. For instance, the task offloading decision and the computational frequency scaling are learned jointly in [10]; in this paper, the latency and energy consumption are minimized by jointly optimizing the task allocation decision and CPU frequency of the mobile user. In [11], the computational speed and transmission power of the mobile devices and the offloading ratio are optimized jointly for reducing energy consumption and latency via dynamic voltage scaling technology. In [12], the transmission power control and offloading decision in the MEC system under a single-user scenario are investigated. As an improvement, the authors in [13] and [14] expand the conclusions in [12] to the multi-user MEC system and combine with the resource allocation to improve the performance of multi-user MEC system by Lyapunov optimization. Similar to [13], the authors in [15] proposed a centralized algorithm for a multi-server multi-user MEC system, which joints the task offloading and resources allocation together. The service delay in MEC is reduced in [16] through virtual machine migration and transmission power control; the users transmit data to cloudlet in a round robin fashion and the service delay is reduced by controlling cloudlets' transmission power. In [17], an integrated framework for computation offloading and interference management in the MEC system is proposed. We have also investigated the transmission power issues in a multi-user interference-aware MEC system [39]; however, in [39], the task offloading decision is not considered. But, the [12] is proposed for single-user MEC system; the algorithm proposed in [16] is mainly for the cloudlets rather than the mobile users; the algorithms proposed in [13], [14], [15], and [17] are effective in multi-user MEC system, but they are centralized; Moreover, for the algorithms introduce above, the feasible region of the transmission power is assumed from 0 to maximum; however, due to the existence of the interference, the transmission power should make the SINR at the MEC server larger than the threshold; thus, the feasible region of the transmission power should be $p \in \{0, [p_{min}, p_{max}]\}$, which will put forward a new challenge for calculating the transmission power of mobile users.

## III. SYSTEM MODEL

There are $N$ mobile users in the network, denoted as $\mathbf{N} = \{1, 2, \ldots, N\}$. The mobile users are served by the BS. In MEC, the FDMA is used in the upper link communication for the task offloading from mobile user to MEC server [18-20]. In FDMA, the available spectrum is divided into $K$ subchannels and the index of these subchannels is $\mathbf{K} = \{1, 2, \ldots, K\}$. In this paper, as [10-15], we only consider the fixed channel allocation, i.e., the channel has been allocated to the mobile users by the BS; in the further version, we will consider introducing the channel allocation into this algorithm, which is interesting and challenging. In the MEC system, each mobile user has a computation-intensive task, which can be calculated locally or offloaded to the MEC server through the BS that deployed proximity to the user. There are two offloading models for mobile users [1]: binary offloading and partial offloading. In this paper, for unifying these two schemas, we introduce the offloading ratio, like that used in [9] and [11], into the offloading decision.

**Definition 1.** The offloading ratio $\lambda \in [0, 1]$ is defined as the ratio of the computation tasks that offloaded to MEC server; when $\lambda = 0$ or $\lambda = 1$, it is binary offloading; when $\lambda \in (0, 1)$, it is partial offloading.

Based on Definition 1, the ratio of the computation task that executed locally is $1 - \lambda$. The transmission power of the mobile user can be adjusted from $p_{min}$ to $p_{max}$. In the wireless networks, the minimum transmission power $p_{min}$ should make the SINR larger than the threshold [18]. Note that when $\lambda = 0$, the computation task will be executed locally, which means $p = 0$; when $\lambda \neq 0$, the task needs to be offloaded to the MEC server and $p \neq 0$; so the feasible region of the transmission power, which takes the task offloading decision and SINR into account, is $p \in \{0, [p_{min}, p_{max}]\}$. This means that the feasible region of transmission power is not continuous anymore, which will put forward a challenge on the transmission power control.

For more clearer, the notations that will be used in this paper are listed in the table below.

TABLE 1.
The notations are used in this paper

| Notations | Meanings |
|---|---|
| $N$ | The number of mobile users in the network |
| $n$ | One of the mobile user |
| $K$ | The number of subchannels |
| $k$ | One of the subchannel |
| $\lambda_n$ | The offloading decision of user $n$ |
| $r_n^k(\boldsymbol{\lambda}, \boldsymbol{p})$ | The data rate of user $n$ when offloads task |
| $w_k$ | The bandwidth of channel $k$ |
| $I_n^k$ | The interference at user $n$ by channel $k$ |
| $p_n$ | The transmission power of user $n$ |
| $\Gamma_n^k$ | The interference of user $n$ by using channel $k$ |
| $T_n$ | The computation task of user $n$ |
| $L_n$ | The size of $T_n$ |
| $c_n$ | The CPU cycles that are needed to process $T_n$ |
| $t_{n,trans}^k$ | The time that is needed to transmit data from user to MEC server |
| $t_n^{cloud}$ | The time that is needed for task executing in MEC server |
| $f_c$ | The CPU capability of MEC server |
| $e_{n,trans}^k$ | The energy consumption that is caused by data transmission from user to MEC server |
| $f_n$ | The CPU capability of user $n$ |
| $O_n^{cloud}$ | The overhead of mobile user $n$ for offloading the computation task to the MEC server |
| $O_n^{local}$ | The overhead of mobile user $n$ for executing the task locally |
| $s_n$ | The strategy of mobile user $n$ |
| $U_n(s_n, s_{-n})$ | The computation overhead of use $n$ that takes the uses in $I_n^k$ into account |

### A. Communication Model

Notice that here we focus on exploring the computation offloading problem under the wireless interference model. This model can well capture user's average aggregate throughput in the cellular communication scenario, and some physical layer channel access schemes are adapted to allow multiple users to share the same spectrum resources simultaneously and efficiently. We define the interference users set of user $n$ in channel $k$ as: *the set of mobile users which use channel $k$ to transmit data to the MEC server, denoted as $I_n^k$*. Therefore, the interference of user $n$ by using channel $k$ to offload the task to MEC server through channel $k$ is: $\Gamma_n^k = \sum_{i \in I_n^k, \lambda_i > 0} p_i G_i$. Based on the Shannon-Hartley formula [24], the transmission rate of mobile user $n$ can be calculated as:

$$r_n^k(\boldsymbol{\lambda}, \boldsymbol{p}) = \omega_k \log_2 \left(1 + \frac{p_n G_n}{N_0 + \Gamma_n^k}\right) \quad (1)$$

where $\omega_k$ is the wireless channel bandwidth; $N_0$ is the power of Gaussian white noise; $G_n$ is the channel gain between mobile user $n$ and BS; $\boldsymbol{\lambda}$ is the set of offloading decisions of the mobile users and $\boldsymbol{\lambda} = \{\lambda_1, \lambda_2, \ldots, \lambda_N\}$; $\boldsymbol{p}$ is the set of transmission powers of the mobile users and $\boldsymbol{p} = \{p_1, p_2, \ldots, p_N\}$. From (1), we can find that the transmission rate of the mobile user $n$ not only relates to its transmission power and offloading decision but also that of the interference users. The transmission of user $n$ is affected by the interference users, and user $n$ also affects the transmission of other users. Due to the interaction between user $n$ and its interference users, the power control and the offloading decision are coupled.

In order to quantify the interference, two kinds of interference sets of user $n$ are defined: (1) the set of users in $I_n^k$ and can affect the data transmission of user $n$ is defined as $I_n^{k,in}$, where $I_n^{k,in} \in I_n^k$; (2) the set of users in $I_n^k$ and can be affected by the data transmission of user $n$ is defined as $I_n^{k,out}$ and $I_n^{k,out} \in I_n^k$. In the following, we will show that these two interference sets are equal. In MEC, when user $i$ decides to offload its task to the MEC server, the SINR of user $i$ at MEC server should be larger than the threshold; the interference of user $i$ comes from the users which use the same channel as user $i$; all the users who use this channel and offload the task to the MEC server can interfere the task offloading of user $i$. The same, user $i$ can interfere all the users who use the same channel as $i$. Thus, we have $I_n^{k,in} = I_n^{k,out} = I_n^k$. Moreover, $I_n^k$ can be got from the MEC server. In the following of this paper, we will use $I_n^k$ to represent $I_n^{k,out}$ and $I_n^{k,in}$.

*B. Computation Model*

For mobile user $n$, there is a computation task $T_n = \{L_n, C_n\}$ which can be calculated locally or offloaded to the MEC server. The $L_n$ is the length of the input data (bits) and $C_n$ is the computation workload (CPU cycles needed for one-bit data).

*B.1 Computation Model of Local Execution*

When $\lambda_n = 0$, the computation task $T_n$ is executed locally, then $p_n = 0$ and $r_n(\lambda, p) = 0$; thus, the latency for calculating this task can be shown as:

$$t_n^{local} = \frac{L_n C_n}{f_n} \qquad (2)$$

where $f_n$ is the CPU cycles per second (i.e., the computation capability) of user $n$ with the upper bound is $f_{max}$. Different mobile users have different computation capabilities and each mobile user can adjust its computation capability from 0 to $f_{max}$. Based on [5] and [13], the energy consumption for calculating the computation task $T_n$ is:

$$e_n^{local} = \kappa_n L_n C_n f_n^2 \qquad (3)$$

In (3), $\kappa_n$ is a constant related to the hardware architecture of mobile device $n$. The dynamic power consumption is proportional to $V_c^2 f_n$, where $V_c$ is the circuit supply voltage. When the operating voltage is low, the CPU frequency is approximately linear proportional to the voltage supply [13, 25-26]. Thus, the energy consumption of one CPU cycle is $\kappa_n f_n^2$. According to (2) and (3), based on the overhead defined in [5], the overhead of local computing is:

$$O_n^{local} = \alpha_{t,n} \frac{L_n C_n}{f_n} + \alpha_{e,n} \kappa_n L_n C_n f_n^2 \qquad (4)$$

where $\alpha_{t,n}, \alpha_{e,n} \in [0,1]$ denote the weights of computational latency and energy for user $n$'s decision making, respectively [5,6]. When user $n$ cares about energy consumption, the user $n$ can set $\alpha_{t,n} = 0$ and $\alpha_{e,n} = 1$; otherwise, when user $n$ is sensitive to delay, then user $n$ can set $\alpha_{t,n} = 1$ and $\alpha_{e,n} = 0$. This model can take both computational latency and energy consumption into decision making at the same time. In practice, the proper weights can be determined by the multi-attribute utility approach in *multi-criteria decision-making theory* [27].

*B.2 Computation Model of MEC server*

When $\lambda_n > 0$, part of or all of the computation tasks are offloaded to the MEC server, where $p_n \in [p_{min}, p_{max}]$ and $r_n(\lambda, p) > 0$. Assuming that the communication channel assigned to mobile user $n$ is $k$, the latency for transmitting the input data to the MEC server can be calculated as:

$$t_{n,trans}^k = \frac{L_n}{r_n^k(\lambda, p)} \qquad (5)$$

If the transmission power of mobile user $n$ for transmitting the computation task to MEC server through channel $k$ is $p_n^k$, the energy consumption of the data transmission is:

$$e_{n,trans}^k = p_n^k t_{n,cloud}^k = \frac{p_n^k L_n}{r_n^k(\lambda, p)} \qquad (6)$$

When the input data is offloaded to the MEC server, the MEC server calculates the computation task based on the input data. According to (2), the latency of task execution in MEC server is:

$$t_n^{cloud} = \frac{L_n C_n}{f_c} \qquad (7)$$

where $f_c$ is the CPU cycle per second of the MEC server.

Therefore, based on (5), (6), and (7), the overhead of cloud computing can be calculated as:

$$O_n^{cloud} = \alpha_{t,n}(t_{n,trans}^k + t_n^{cloud}) + \alpha_{e,n} e_{n,trans}^k \qquad (8)$$

According to the offloading decision ratio $\lambda_n$, the length of the input data that calculated locally is $(1 - \lambda_n)L_n$ and the length of the input data that offloaded to the MEC server is $\lambda_n L_n$ [9][11]. Thus, the whole computation overhead of user $n$, which takes the overheads of local computing and cloud computing into account, can be calculated as:

$$O_n = \alpha_{t,n}\left(\frac{\lambda_n L_n}{r_n^k(\lambda,p)} + \frac{\lambda_n L_n C_n}{f_c} + \frac{(1-\lambda_n)L_n C_n}{f_n}\right)$$
$$+ \alpha_{e,n}\left(\frac{\lambda_n p_n^k L_n}{r_n^k(\lambda,p)} + (1-\lambda_n)\kappa_n L_n C_n f_n^2\right) \qquad (9)$$

## IV. JOINT TASK OFFLOADING AND RESOURCES ALLOCATION GAME

Based on Section 1.3, the problem needs to be solved in this paper can be shown in **P1**:

$$\textbf{P1}: \min_{n \in N} O_n(\lambda_n, p_n, f_n)$$

$$s.t. \quad 0 \leq \lambda_n \leq 1, n \in \mathbf{N} \qquad (a)$$
$$p_n \in \begin{cases} [p_{min}, p_{max}], \lambda_n > 0, n \in \mathbf{N} \\ \{0\}, \qquad \lambda_n = 0, n \in \mathbf{N} \end{cases} \quad (b) \qquad (10)$$
$$0 \leq f_n \leq f_{max}, n \in \mathbf{N} \qquad (c)$$

According to the **P1** shown in (10), we define the multi-user game as: $G = (N, \{S_n\}_{n \in N}, \{O_n\}_{n \in N})$, where $S_n = \lambda_n \otimes p_n \otimes f_n$ is the strategy space of mobile user $n$; $\lambda_n, p_n$, and $f_n$ are the strategy spaces of offloading decision, transmission power, and CPU capability of user $n$, respectively, i.e., $\lambda_n \in \lambda_n, p_n \in p_n$, and $f_n \in f_n$. For each strategy of user $n$, $s_n \in S_n$ and $s_n = \{\lambda_n, p_n, f_n\}$. Moreover, $S = \{S_1, S_2, \ldots, S_N\}$ is the strategy space set of all mobile users. The utility function of mobile user $n$ is $O_n(s_n, s_{-n})$, in which $s_{-n} = (s_1, s_2, \ldots, s_{n-1}, s_{n+1}, \ldots, s_N)$ is the set of strategies of all the others mobile users except user $n$. So the game $G$ can be stated as that for given $s_{-n}$ (can be gotten from the BS), the mobile user $n$ would like to choose a proper strategy $s_n = \{\lambda_n, p_n, f_n\}$ to minimize its computation overhead, i.e.,

$$G: \min_{s_n \in S_n} O_n(s_n, s_{-n}), \forall n \in \mathbf{N} \quad (11)$$

However, as discussed in (1), the offloading decision of user $n$ can affect the performances of users in $I_n^k$; moreover, according to the conclusions in [19], [20], and [28], the approach which defines the individual computation overhead as the utility of each user is selfish. So, the local altruistic behaviors among neighboring users, which are motivated by local cooperation in biographical systems [29-30], are introduced into the utility function construction [19-20, 28]. In game $G$, when the strategy of user $n$ changes, the computation overheads of the users in $I_n^k$ are affected; therefore, in the new overhead function, it is better to take the computation overhead of the users in $I_n^k$ into account [19][20][28], which can be expressed as:

$$U_n(s_n, s_{-n}) = O_n(s_n, s_{-n}) + \sum_{i \in I_n^k} O_i(s_{i,s_n}, s_{-i}) \quad (12)$$

where $s_{i,s_n}$ means the strategy of user $i \in I_n^k$ when the strategy of user $n$ is $s_n$. In (12), the first term in the right hand is the computation overhead of user $n$; the second term is the aggregated computation overhead of the users in $I_n^k$. Then the new local altruistic game $G'$ is:

$$G': \min_{s_n \in S_n} U_n(s_n, s_{-n}), \forall n \in \mathbf{N} \quad (13)$$

Based on the definition of game $G'$ in (13), we define the NE of game $G'$ as follows.

**Definition 2.** A strategy profile $\mathbf{s}^* = \{s_1^*, s_2^*, \ldots, s_N^*\}$ is a NE of game $G'$, if at the NE point $\mathbf{s}^*$, no mobile user can reduce their computation overhead by changing the strategy unilaterally. The mathematic formula expression is: for $\forall s_n \in S_n$ and $\forall n \in \mathbf{N}$, $U_n(s_n^*, s_{-n}^*) \leq U_n(s_n, s_{-n}^*)$ holds.

At the NE point, each user cannot find a better strategy than the current one when the other users do not change their strategies. This property is important to the distributed issues, since each mobile user minimizes their own overhead based on their own interests, which will reduce the information exchange between mobile users and the MEC server.

*A. The Existence and property of Nash Equilibrium*

Since the computation overhead $U_n(s_n, s_{-n})$ and the strategy space of transmission power $p_n$ are not continuous, so we introduce the potential game into the proving of the existence and uniqueness of NE. For proving that $G'$ is a potential game, first, we define the potential game in Definition 3.

**Definition 3** [31][32]. A game is an exact potential game if there exists a potential function $\Phi$ such that for $\forall n \in \mathbf{N}$ and $\forall s_n, s_n', s_{-n} \in S_n$, the following conditions hold:

$$U_n(s_n', s_{-n}) - U_n(s_n, s_{-n}) = \Phi(s_n', s_{-n}) - \Phi(s_n, s_{-n}) \quad (14)$$

*Remark 1:* For the exact potential game, if any mobile user changes its strategy (i.e., from $s_n$ to $s_n'$), the variation in the overhead function equals to that in the potential function. The important property of the potential game is that there always exists a NE and the asynchronous better response update process must be finite and leads to a NE [31-32].

**Theorem 1.** The game $G'$ shown in (13) is an exact potential game; and the game $G'$ always has a Nash equilibrium and finite improvement property.

*Proof.* Based on the conclusions in [19], [20], and [28], we define the potential function as:

$$\begin{aligned}\Phi(s_n, s_{-n}) &= \sum_{n \in N} O_n(s_n, s_{-n}) \\ &= O_n(s_n, s_{-n}) + \sum_{i \in I_n^k} O_i(s_{i,s_n}, s_{-i}) + \\ & \quad \sum_{j \in N \setminus I_n^k \setminus \{n\}} O_j(s_{j,s_n}, s_{-j}) \end{aligned} \quad (15)$$

The (15) can be divided into three terms. The first term in the right hand is the computation overhead of mobile user $n$; the second term represents the summary of the overheads of the users in $I_n^k$; the third term is the computation overhead of the rest mobile users in the network. Since there are three variables in $s_n$, so there will be $\sum_{i=1}^{3} C_3^i = 7$ different changing models of $s_n$. If we calculate the variation of the computation overhead function and the potential function based on all the changing models of $s_n$, it will be complex. However, the variables in $s_n$ can be divided into two different groups: 1) $\lambda_n$ and $p_n$; these two variables can not only affect the overhead of user $n$ but also the users in $I_n^k$, denoted as $par$I; 2) $f_n$; when this variable changes, only the overhead of user $n$ is affected, denoted as $par$II.

When $par$I changes, i.e., $\lambda_n$ or $p_n$ changes or both of these two variables change, no matter the $par$II changes or not, the overheads of user $n$ and the users in $I_n^k$ change; the computation overheads of the rest users (i.e., the users in $N \setminus I_n^k \setminus \{n\}$) do not change, because the strategy changing of user $n$ cannot affect the mobile users which are not in $I_n^k$. The deviation of the computation overhead when the user $n$'s strategy changes from $s_n$ to $s_n'$ can be calculated:

$$\begin{aligned}U_n(s_n', s_{-n}) &- U_n(s_n, s_{-n}) \\ &= O_n(s_n', s_{-n}) - O_n(s_n, s_{-n}) + \sum_{i \in I_n^k} O_i(s_{i,s_n'}, s_{-i}) \\ & \quad - \sum_{i \in I_n^k} O_i(s_{i,s_n}, s_{-i})\end{aligned} \quad (16)$$

According to (15), the variation of the potential function is:

$$\begin{aligned}\Phi(s_n', s_{-n}) &- \Phi(s_n, s_{-n}) \\ &= O_n(s_n, s_{-n}) - O_n(s_n', s_{-n}) + \sum_{i \in I_n^k} O_i(s_{i,s_n'}, s_{-i}) \\ & \quad - \sum_{i \in I_n^k} O_i(s_{i,s_n}, s_{-i}) + \sum_{j \in N \setminus I_n^k \setminus \{n\}} O_j(s_{j,s_n'}, s_{-j}) \\ & \quad - \sum_{j \in N \setminus I_n^k \setminus \{n\}} O_j(s_{j,s_n}, s_{-j})\end{aligned} \quad (17)$$

As shown in (17), $\sum_{j \in N \setminus I_n^k \setminus \{n\}} O_j(s_{j,s_n'}, s_{-j}) - \sum_{j \in N \setminus I_n^k \setminus \{n\}} O_j(s_{j,s_n}, s_{-j}) = 0$, so $U_n(s_n', s_{-n}) - U_n(s_n, s_{-n}) = \Phi(s_n', s_{-n}) - \Phi(s_n, s_{-n})$ holds.

When the $par$II changes and the $par$I does not change, the computation overhead of the other mobile users except user $n$ keeps constant. So based on (16), $\sum_{i \in I_n^k} O_i(s_{i,s_n'}, s_{-i}) - \sum_{i \in I_n^k} O_i(s_{i,s_n}, s_{-i}) = 0$ and $\sum_{j \in N \setminus I_n^k \setminus \{n\}} O_j(s_{j,s_n'}, s_{-j}) - \sum_{j \in N \setminus I_n^k \setminus \{n\}} O_j(s_{j,s_n}, s_{-j}) = 0$. Thus, the conclusion in (14) holds. Therefore, the $G'$ is an exact potential game.

Based on the conclusions in [5], [31], [32], and [34], since the game $G'$ is an exact potential game, so it always has a Nash equilibrium and the finite improvement property. ∎

## B. The best response strategy

For the game which the existence of NE is guaranteed, the best-response dynamic always converges to a NE [33-34], so it is applied in this algorithm to reach the NE of game $G'$. In the best response strategy, each mobile user calculates the best response of the variables in $s$ according to the information that gotten from BS, i.e., for given $s_{-n}$, user $n$ calculates the best response of $\lambda_n$, $p_n$, and $f_n$ based on (12).

**Corollary 1.** For $\forall p_n \in \boldsymbol{p}_n$ and $\forall f_n \in \boldsymbol{f}_n$, the best response of $\lambda_n$ will be $\lambda_n = 0$ or $\lambda_n = 1$.

*Proof.* Based on $O_n(s_n, s_{-n})$, the computation overhead function of game $G'$ can be expressed as:

$$U_n(s_n, s_{-n}) = \alpha_{t,n}\left(\frac{\lambda_n L_n}{r_n^k(\boldsymbol{\lambda},\boldsymbol{p})} + \frac{\lambda_n L_n C_n}{f_c} + \frac{(1-\lambda_n)L_n C_n}{f_n}\right)$$
$$+\alpha_{e,n}\left(\frac{\lambda_n p_n^k L_n}{r_n^k(\boldsymbol{\lambda},\boldsymbol{p})} + (1-\lambda_n)\kappa_n L_n C_n f_n^2\right)$$
$$+\sum_{i\in I_n^k}\left(\alpha_{t,i}\left(\frac{\lambda_i L_i}{r_i^k(\boldsymbol{\lambda},\boldsymbol{p})} + \frac{\lambda_i L_i C_i}{f_c} + \frac{(1-\lambda_i)L_i C_i}{f_i}\right)\right.$$
$$\left.+\alpha_{e,i}\left(\frac{\lambda_i p_i^k L_i}{r_i^k(\boldsymbol{\lambda},\boldsymbol{p})} + (1-\lambda_i)\kappa_i L_i C_i f_i^2\right)\right) \quad (18)$$

Since when $\lambda_n = 0$, $p_n = 0$, when $\lambda_n \neq 0$, $p_n \in [p_{min}, p_{max}]$, so the $U_n(s_n, s_{-n})$ is not continuous. For solving this issues, the value of $\lambda_n$ is divided into two parts: 1) $\lambda_n \in [\varepsilon, 1]$, where $\varepsilon$ is small enough and can near to 0 arbitrarily, then $p_n \in [p_{min}, p_{max}]$; 2) $\lambda_n = 0$, then $p_n = 0$. When $\lambda_n \in [\varepsilon, 1]$ and $p_n \neq 0$, the third term in (18) has no relation with $\lambda_n$, because according to (1), when $\lambda_n \in [\varepsilon, 1] > 0$, the user $n$ will interfere the users in $I_n^k$ no matter what the value of $\lambda_n$ is. Obviously, the $O_n(s_n, s_{-n})$ is a linear function on $\lambda_n$; so the extreme value of $O_n(s_n, s_{-n})$ will be gotten when $\lambda_n = \varepsilon$ or $\lambda_n = 1$. When $\alpha_{t,n}\left(\frac{L_n}{r_n^k(\boldsymbol{\lambda},\boldsymbol{p})} + \frac{L_n C_n}{f_c}\right) + \alpha_{e,n}\frac{p_n^k L_n}{r_n^k(\boldsymbol{\lambda},\boldsymbol{p})} > \alpha_{t,n}\frac{L_n C_n}{f_n} + \alpha_{e,n}\kappa_n L_n C_n f_n^2$, the $O_n(s_n, s_{-n})$ is an increasing function and the minimum computation overhead can be gotten when $\lambda_n = \varepsilon$. Thus, when $\varepsilon \to 0$, the $U_n(s_n, s_{-n})$ can be calculated as:

$$\lim_{\varepsilon \to 0} U_n(s_n, s_{-n}) = \alpha_{t,n}\frac{L_n C_n}{f_n} + \alpha_{e,n}\kappa_n L_n C_n f_n^2$$
$$+\sum_{i\in I_n^k}\left(\alpha_{t,i}\left(\frac{\lambda_i L_i}{r_{i,p_n\neq 0}^k(\boldsymbol{\lambda},\boldsymbol{p})} + \frac{\lambda_i L_i C_i}{f_c} + \frac{(1-\lambda_i)L_i C_i}{f_i}\right)\right.$$
$$\left.+\alpha_{e,i}\left(\frac{\lambda_i p_i^k L_i}{r_{i,p_n\neq 0}^k(\boldsymbol{\lambda},\boldsymbol{p})} + (1-\lambda_i)\kappa_i L_i C_i f_i^2\right)\right) \quad (19)$$

When $\alpha_{t,n}\left(\frac{L_n}{r_n^k(\boldsymbol{\lambda},\boldsymbol{p})} + \frac{L_n C_n}{f_c}\right) + \alpha_{e,n}\frac{p_n^k L_n}{r_n^k(\boldsymbol{\lambda},\boldsymbol{p})} < \alpha_{t,n}\frac{L_n C_n}{f_n} + \alpha_{e,n}\kappa_n L_n C_n f_n^2$, the $O_n(s_n, s_{-n})$ is a decreasing function; so the minimum overhead will be gotten when $\lambda_n = 1$, which is:

$$U_n(s_n, s_{-n})|_{\lambda_n=1} = \alpha_{t,n}\left(\frac{L_n}{r_n^k(\boldsymbol{\lambda},\boldsymbol{p})} + \frac{L_n C_n}{f_c}\right) + \alpha_{e,n}\frac{p_n^k L_n}{r_n^k(\boldsymbol{\lambda},\boldsymbol{p})}$$
$$+\sum_{i\in I_n^k}\left(\alpha_{t,i}\left(\frac{\lambda_i L_i}{r_{i,p_n\neq 0}^k(\boldsymbol{\lambda},\boldsymbol{p})} + \frac{\lambda_i L_i C_i}{f_c} + \frac{(1-\lambda_i)L_i C_i}{f_i}\right)\right.$$
$$\left.+\alpha_{e,i}\left(\frac{\lambda_i p_i^k L_i}{r_{i,p_n\neq 0}^k(\boldsymbol{\lambda},\boldsymbol{p})} + (1-\lambda_i)\kappa_i L_i C_i f_i^2\right)\right) \quad (20)$$

When $\lambda_n = 0$, the computation overhead shown in (18) can be calculated as:

$$U_n(s_n, s_{-n})|_{\lambda_n=0} = \alpha_{t,n}\frac{L_n C_n}{f_n} + \alpha_{e,n}\kappa_n L_n C_n f_n^2$$
$$+\sum_{i\in I_n^k}\left(\alpha_{t,i}\left(\frac{\lambda_i L_i}{r_{i,p_n=0}^k(\boldsymbol{\lambda},\boldsymbol{p})} + \frac{\lambda_i L_i C_i}{f_c} + \frac{(1-\lambda_i)L_i C_i}{f_i}\right)\right.$$
$$\left.+\alpha_{e,i}\left(\frac{\lambda_i p_i^k L_i}{r_{i,p_n=0}^k(\boldsymbol{\lambda},\boldsymbol{p})} + (1-\lambda_i)\kappa_i L_i C_i f_i^2\right)\right) \quad (21)$$

According to (1), we have $r_{i,p_n\neq 0}^k(\boldsymbol{\lambda},\boldsymbol{p}) = \omega_k \log_2\left(1+\frac{p_i^k G_i}{N_0+\sum_{j\in I_i^k}p_j^k G_j+p_n^k G_n}\right)$ and $r_{i,p_n=0}^k(\boldsymbol{\lambda},\boldsymbol{p}) = \omega_k \log_2\left(1+\frac{p_i^k G_i}{N_0+\sum_{j\in I_i^k}p_j^k G_j}\right)$, so $r_{i,p_n\neq 0}^k(\boldsymbol{\lambda},\boldsymbol{p}) < r_{i,p_n=0}^k(\boldsymbol{\lambda},\boldsymbol{p})$; therefore, based on (19) and (21), $\lim_{\varepsilon\to 0} U_n(s_n, s_{-n}) > U_n(s_n, s_{-n})|_{\lambda_n=0}$. So in this case, the best response of $\lambda_n$ is $\lambda_n = 0$.

When $\alpha_{t,n}\left(\frac{L_n}{r_n^k(\boldsymbol{\lambda},\boldsymbol{p})} + \frac{L_n C_n}{f_c}\right) + \alpha_{e,n}\frac{p_n^k L_n}{r_n^k(\boldsymbol{\lambda},\boldsymbol{p})} < \alpha_{t,n}\frac{L_n C_n}{f_n} + \alpha_{e,n}\kappa_n L_n C_n f_n^2$, the overhead function shown in (18) is a decreasing function. On the one hand, the first two terms in (20) are smaller than that in (21); on the other hand, the third term in (20) is larger than that in (21) due to $r_{i,p_n\neq 0}^k(\boldsymbol{\lambda},\boldsymbol{p}) < r_{i,p_n=0}^k(\boldsymbol{\lambda},\boldsymbol{p})$. This means that $U_n(s_n, s_{-n})|_{\lambda_n=1}$ may larger or smaller than $U_n(s_n, s_{-n})|_{\lambda_n=0}$. Therefore, in this case, the best response of $\lambda_n$ is $\lambda_n = 1$ or $\lambda_n = 0$ according to the values of $U_n(s_n, s_{-n})|_{\lambda_n=1}$ and $U_n(s_n, s_{-n})|_{\lambda_n=0}$. Thus, Corollary 1 holds. ∎

*Remark 2*: In the proof of Corollary 1, when $\alpha_{t,n}\left(\frac{L_n}{r_n^k(\boldsymbol{\lambda},\boldsymbol{p})} + \frac{L_n C_n}{f_c}\right) + \alpha_{e,n}\frac{p_n^k L_n}{r_n^k(\boldsymbol{\lambda},\boldsymbol{p})} > \alpha_{t,n}\frac{L_n C_n}{f_n} + \alpha_{e,n}\kappa_n L_n C_n f_n^2$, the value of $\lambda_n$ is $\lambda_n = 0$. This is easy to be understood; since in this case, the task is calculated locally costs less resource than that in the MEC server. However, when $\alpha_{t,n}\left(\frac{L_n}{r_n^k(\boldsymbol{\lambda},\boldsymbol{p})} + \frac{L_n C_n}{f_c}\right) + \alpha_{e,n}\frac{p_n^k L_n}{r_n^k(\boldsymbol{\lambda},\boldsymbol{p})} < \alpha_{t,n}\frac{L_n C_n}{f_n} + \alpha_{e,n}\kappa_n L_n C_n f_n^2$, the best response of $\lambda_n$ is $\lambda_n = 1$ or $\lambda_n = 0$; this means that even the user n's computation overhead of cloud computing is smaller than that of local computing, considering the overheads of the users in $I_n^k$, the user n may not offloads its task to MEC server.

**Corollary 2.** The best response of the CPU capability $f_n$ is $f_n = (\alpha_{t,n}/2\alpha_{e,n}\kappa_n)^{1/3}$ or $f_n = 0$.

*Proof.* When $\lambda_n \neq 0$, $f_n = 0$; only when $\lambda_n = 0$, $f_n \in (0, f_{max}]$. Thus, when $\lambda_n = 1$, the best response of $f_n$ is $f_n = 0$. When $\lambda_n = 0$, the computation overhead function is shown in (21). The first derivative of (21) on $f_n$ is: $U_n'(s_n, s_{-n})|_{f_n} = 2\alpha_{e,n}\kappa_n L_n C_n f_n - \alpha_{t,n}\frac{L_n C_n}{f_n^2}$. When $U_n'(s_n, s_{-n})|_{f_n} = 0$, $f_n = (\alpha_{t,n}/2\alpha_{e,n}\kappa_n)^{1/3}$; moreover, since $U_n''(s_n, s_{-n})|_{f_n} = 2\alpha_{e,n}\kappa_n L_n C_n + 2\alpha_{t,n}\frac{L_n C_n}{f_n^3} > 0$, so when $f_n = (\alpha_{t,n}/2\alpha_{e,n}\kappa_n)^{1/3}$, the $U_n(s_n, s_{-n})$ can get the minimum value. Thus, the best response of the CPU capability $f_n$ is $f_n = (\alpha_{t,n}/2\alpha_{e,n}\kappa_n)^{1/3}$ or $f_n = 0$. ∎

**Corollary 3.** For $\forall \lambda_n \in \{0,1\}$, the best response of $p_n$ exists and $p_n = 0$ or $p_n = \bar{p}_n$, where $\bar{p}_n = \arg\min_{n\in N}\{U_n(s_{n|p_n=p_{min}}, s_{-n}), U_n(s_{n|p_n=p_n'}, s_{-n}), U_n(s_{n|p_n=p_{max}}, s_{-n})\}$.

*Proof.* Since the best response of offloading decision is $\lambda_n \in \{0,1\}$ and when $\lambda_n = 0$, $p_n = 0$, so the best response of $p_n$ when $\lambda_n = 0$ is $p_n = 0$. When $\lambda_n = 1$, the task is offloaded to the MEC server. Since the delay and the energy consumption when the task is executed in MEC server have no relation with the transmission power, so we only consider the latency and the energy consumption caused by the data transmission. According to (18), the transmission overhead can be calculated as $U_n(s_n, s_{-n})_{trans} = \alpha_{t,n}\frac{L_n}{r_n^k(\lambda,\boldsymbol{p})} + \alpha_{e,n}\frac{p_n^k L_n}{r_n^k(\lambda,\boldsymbol{p})} + \sum_{i\in I_n^k}U_i(s_i, s_{-i})$; considering (1), the overhead of transmission can be written as:

$$U_n(s_n, s_{-n})_{trans} = a_n\frac{\alpha_{t,n}+\alpha_{e,n}p_n^k}{\ln\left(1+\left(p_n^k G_n/\Gamma_n^k\right)\right)} + \sum_{i\in I_n^k}\frac{b_i}{\ln\left(1+\left(p_i^k G_i/N_0+p_n^k G_n+\sum_{j\in I_i^{k,in}\{n\}}p_j^k G_j\right)\right)} \quad (22)$$

where $a_n = L_n \ln 2/\omega_k$, $b_i = \lambda_i L_n(\alpha_{t,i}+\alpha_{e,i}p_i^k)\ln 2/\omega_k$. The extreme value of (22) can be gotten when $U'_n(s_n, s_{-n})_{trans}|_{p_n^k} = 0$. Since $U'_n(s_n, s_{-n})_{trans}|_{p_n^k} = 0$ is a transcendental equation, so the analytical solution of $U'_n(s_n, s_{-n})_{trans}|_{p_n^k} = 0$ does not exist. The numerical solution of $U'_n(s_n, s_{-n})_{trans}|_{p_n^k} = 0$ can be calculated by Newton Method. Let $p'_n$ be the solution of $U'_n(s_n, s_{-n})_{trans}|_{p_n^k} = 0$; since the solution of $U'_n(s_n, s_{-n})_{trans}|_{p_n^k} = 0$ may not single, so we define the $p'_n$ as: $p'_n = \arg\{U'_n(s_n, s_{-n})_{trans}|_{p_n^k} = 0\}$. Therefore, if $p'_n = \emptyset$, the $U_n(s_n, s_{-n})_{trans}$ is a monotone function with $p_n \in [p_{min}, p_{max}]$, then the best response of $p_n$ will be $p_n = p_{min}$ or $p_n = p_{max}$. If $p'_n \neq \emptyset$, the extreme value of (22) will be gotten at $p'_n$; therefore, if $p'_n \in [p_{min}, p_{max}]$, it means that the extreme value of (22) exists in the feasible region $[p_{min}, p_{max}]$; if $p'_n \notin [p_{min}, p_{max}]$, the extreme value of (22) can be gotten when $p'_n = p_{min}$ or $p'_n = p_{max}$. Thus, the best response of the transmission power $p_n$ can be calculated as $\bar{p}_n = \arg\min_{n\in N}\{U_n(s_{n|p_n=p_{min}}, s_{-n}), U_n(s_{n|p_n=p'_n}, s_{-n}), U_n(s_{n|p_n=p_{max}}, s_{-n})\}$. Therefore, the best response of $p_n$ is $p_n = 0$ or $p_n = \bar{p}_n$. ∎

**Corollary 4.** The best response of $\lambda_n$ is decided by the interference from the mobile users in $I_n^k$ and $I_i^k$, where $I_i^k$ is the set of the interference users which can affect the data transmission of user $i \in I_n^k$.

*Proof.* Since the best response of $\lambda_n$ is $\lambda_n = 0$ or $\lambda_n = 1$, so the best response of $U_n(s_n, s_{-n})$ can be expressed as:

$$U_n^*(s_n, s_{-n}) = \begin{cases} U_1^* = U_n^*(s_n|_{p_n=\bar{p}_n}, s_{-n})_{trans} + U_n^*(s_n|_{f_n=0}, s_{-n})_{cloud} & \lambda_n = 1 \\ U_0^* = U_n^*(s_n|_{p_n=0}, s_{-n})_{trans} + U_n^*\left(s_n|_{f_n=\left(\frac{\alpha_{t,n}}{2\alpha_{e,n}\kappa_n}\right)^{\frac{1}{3}}}, s_{-n}\right)_{local} & \lambda_n = 0 \end{cases} \quad (23)$$

where $U_n^*(s_n|_{p_n=\bar{p}_n}, s_{-n})_{trans}$ and $U_n^*(s_n|_{p_n=0}, s_{-n})_{trans}$ are the best response of $U_n(s_n, s_{-n})_{trans}$, which can be gotten when $p_n = \bar{p}_n$ and $p_n = 0$ in (22), respectively; $U_n^*(s_n|_{f_n=0}, s_{-n})_{cloud} = \alpha_{t,n}L_n C_n/f_c$ is the best response of the computation overhead in MEC server; $U_n^*\left(s_n|_{f_n=\left(\frac{\alpha_{t,n}}{2\alpha_{e,n}\kappa_n}\right)^{1/3}}, s_{-n}\right)_{local}$ is the best response of the computation overhead in mobile user n.

For the best response strategy, if $U_1^* < U_0^*$, then $\lambda_n = 1$; otherwise, if $U_1^* > U_0^*$, then $\lambda_n = 0$. From (23), we can conclude that the values of $U_1^*$ and $U_0^*$ relate to the best response of $\lambda_n$, $p_n$, $f_n$, and the interference from the interference users (the users in $I_n^k$ and $I_i^k$); moreover, the value of $(\alpha_{t,n}/2\alpha_{e,n}\kappa_n)^{1/3}$ is constant and the best response of $p_n$ is determined by the interference from the users in $I_n^k$. So we can conclude that $U_1^*$ and $U_0^*$ will be decided by the interference from the interference users. This means that the best response of $\lambda_n$ is also decided by the interference from the interference users. Thus, Corollary 4 holds. ∎

*Remark 3:* The Corollary 4 is easy to be understood since for the mobile user $n$, if the interference from the interference users is high, then calculating the task locally is beneficial; otherwise, offloading the task to the MEC server is the best choice. The Corollary 4 does not mean that $\lambda_n$ and $p_n$ are independent, since both $\lambda_n$ and $p_n$ are affected by the interference caused by the data transmission of the interference users. The Corollary 4 indicates that minimizing the interference (e.g., by transmission power control or offloading decision) is one possible approach to improve the performance of MEC.

### C. Process of the proposed algorithm

Since we have proved the existence of the NE, so the main idea of this algorithm is to let the mobile users improve their strategies at each time slot and reach the NE at the end based on the best-response dynamic.

At the beginning of each time slot, the BS allocates channels to the mobile users and measures the channel interference [5][18][19][20]; the measured channel interference will be sent to user $n$. Then, user $n$ calculates the best responses of $\lambda_n$, $p_n$, and $f_n$, and updates its strategy based on these best responses. First, the user $n$ calculates the values of $\bar{p}_n$ and $(\alpha_{t,n}/2\alpha_{e,n}\kappa_n)^{1/3}$ according to the Corollary 2 and Corollary 3; when these two values are gotten, the best response of $U_1^*$ and $U_0^*$ can be calculated. The offloading decision $\lambda_n$ is decided based on the rules as follows: *If* $U_1^* < U_0^*$, *then* $\lambda_n = 1$; *otherwise, if* $U_1^* > U_0^*$, $\lambda_n = 0$. Once the offloading decision is determined, the best response of the transmission power $p_n$ and the CPU capability $f_n$ can be decided based on: *If* $\lambda_n = 0$, *then* $p_n = 0$ *and* $f_n = (\alpha_{t,n}/2\alpha_{e,n}\kappa_n)^{1/3}$; *otherwise, if* $\lambda_n = 1$, *then* $p_n = \bar{p}_n$ *and* $f_n = 0$. This process will be executed repeatedly until the NE is reached.

| **Algorithm 1**: Joint offloading decision and resources allocation |
|---|
| 1. **initialization:** |
| 2. each mobile user $n$ chooses the offloading decision, the transmission power, and CPU capability as: $\lambda_n = 1$, $p_n = p_{max}$, and $f_n = 0$; |
| 3. **end** |
| 4. **repeat** for each user and each decision slot in parallel: |
| 5. send the pilot signal on the chosen communication channel to the base station; |
| 6. receive the channel interference from BS; |
| 7. calculate the best response of the transmission power $\bar{p}_n$ and the CPU capability $(\alpha_{t,n}/2\alpha_{e,n}\kappa_n)^{1/3}$; |
| 8. calculate the $U_1^*$ and $U_0^*$ based on $\bar{p}_n$ and $(\alpha_{t,n}/2\alpha_{e,n}\kappa_n)^{1/3}$; |
| 9. **if** $U_1^* > U_0^*$ |
| 10. $\lambda_n = 0$; |
| 11. **else if** $U_1^* < U_0^*$ |
| 12. $\lambda_n = 1$; |
| 13. **end if** |
| 14. **if** $\lambda_n = 0$ |
| 15. $p_n = 0$ and $f_n = (\alpha_{t,n}/2\alpha_{e,n}\kappa_n)^{1/3}$; |
| 16. **else if** $\lambda_n = 1$ |

17.     $p_n = \bar{p}_n$ and $f_n = 0$;
18.     **end if**
19.     **repeat until** NE is meet.

## V. PERFORMANCE ANALYSIS

### A. Convergence and Computational Complexity

In [31] and [32], the authors have proved that for any potential game, the best response dynamics always converge to a pure Nash Equilibrium. Since the game $G'$ is an exact potential game, so the algorithm proposed in this paper is convergent.

**Corollary 5.** Finding the NE of game $G'$ by the best response approach is PLS-complete and the computational complexity is $O[Cn\log^3(n)]$.

*Proof.* In [31] and [32], the authors have proved that for the potential game which applies the best response approach in the process of reaching NE, if the best response can be computed in polynomial time, then this problem is PLS (Polynomial Local Search) complete. In game $G'$, there are three best responses that need to be calculated in each time slot, which are $\lambda_n$, $p_n$, and $f_n$. The computational complexity for calculating the best response of $f_n$ is $O(1)$, since the best response of $f_n$ is constant. For calculating the best response of $p_n$, the Newton Method is applied. In [35], the authors have proved that for $f(x)$, the computation complexity of the Newton Method is $O[\log(n)F(n)]$, where $F(n)$ is the computation cost of $f(x)/f'(x)$. Moreover, in game $G'$, $f(x) = U_n(s_n, s_{-n})$. Since $U_n(s_n, s_{-n})$ and $U'_n(s_n, s_{-n})$ are all polynomial, so the $U_n(s_n, s_{-n})/U'_n(s_n, s_{-n})$ is also polynomial. This means that the computation of $F(n)$ can be completed in polynomial time. Thus, the calculation of the best response of $p_n$ can be finished in polynomial time. Based on the Corollary 4, the best response of $\lambda_n$ relates to the calculation of $\bar{p}_n$ and $(\alpha_{t,n}/2\alpha_{e,n}\kappa_n)^{1/3}$; therefore, the computation of $\lambda_n$ also can be completed at polynomial time.

For game $G'$, at each time slot, the computational complexity relates to the best response calculation of $\lambda_n$, $p_n$, and $f_n$. As shown in the Corollary 5, the complexity of the best response computation of $f_n$ is $O(1)$, which is much simpler than that of $p_n$ and $\lambda_n$, and can be ignored. Moreover, when the best response of $p_n$ and $f_n$ are gotten, the complexity of the best response calculation of $\lambda_n$ is also simple, which is shown in (23). Therefore, the main computational complexity of this algorithm is caused by the calculation of the best response of $p_n$, i.e., the Newton Method. So the computational complexity of game $G'$ is $O[\log(n)F(n)]$ in one-time slot, where $F(n)$ is the computational cost of $U_n(s_n, s_{-n})/U'_n(s_n, s_{-n})$. According to (18), the computational cost of $F(n)$ is $O[n\log^2(n)]$. Therefore, the computational complexity of this algorithm in one-time slot is $O[n\log^3(n)]$. Assuming that the algorithm needs $C$ rounds iteration for reaching NE, then the computation complexity of the proposed algorithm is [5]: $O[Cn\log^3(n)]$. This demonstrates that the algorithm can be completed at polynomial time. Thus, the Corollary 5 holds. ∎

### B. Properties of the game

In this section, we learn the PoA of the computation overhead of the whole network, i.e., $\sum_{n\in N, s_n\in s} U_n$. Based on the conclusion in [36], the PoA is defined as:

$$PoA = \frac{\sum_{n\in N, s_n\in s} U_n(\tilde{s})}{\sum_{n\in N, s_n\in s} \bar{U}_n(s^*)} \quad (24)$$

where $\tilde{s}$ is the NE of game $G'$, $s^*$ is the centralized optimal solution for all the mobile users, $U_n$ is the overhead of user $n$ by the game theory based algorithm; $\bar{U}_n$ is the overhead by the centralized algorithm. The PoA represents the efficiency ratio of the worst-case of NE over the centralized optimal solutions. For the MEC, the smaller the PoA is, the better the performance is [5].

**Corollary 6.** For game $G'$, the PoA of the network computation overhead satisfies that:

$$1 \le PoA \le \frac{\sum_{n=1,\lambda_n=1}^{N}(U_{n,c}^{max}+\sum_{i\in I_n^k} U_i)+\sum_{n=1,\lambda_n=0}^{N}(U_n^{max}+\sum_{i\in I_n^k} U_i)}{\sum_{n=1,\lambda_n=1}^{N}(\bar{U}_{n,c}^{min}+\sum_{i\in I_n^k} U_i)+\sum_{n=1,\lambda_n=0}^{N}(\bar{U}_n^{min}+\sum_{i\in I_n^k} U_i)} \quad (25)$$

where $U_{n,c}^{max} = \frac{(\alpha_t+\alpha_e p_n)s_n}{w_k \log_2\left(1+\frac{p_n G_n}{N_0+\sum_{j\in I_n^k, \lambda_j=1} p'_{max} G_j}\right)} + \alpha_t t_{n,c} + \alpha_e e_{n,c}$, $U_n^{max} = \frac{\alpha_t c_n}{f_n^{max}} + \alpha_e \kappa_n c_n (f_n^{max})^2$, $\bar{U}_{n,c}^{max} = \frac{(\alpha_t+\alpha_e p_n)s_n}{w_k \log_2\left(1+\frac{p_n G_n}{N_0}\right)} + \alpha_t t_{n,c} + \alpha_e e_{n,c}$, $\bar{U}_n^{max} = \frac{\alpha_t c_n}{f_n^*} + \alpha_e \kappa_n c_n (f_n^*)^2$, and $p'_{max} \triangleq \max\{p_j, j\in I_n^k\}$. $p'_{max}$ is the maximum transmission power of all mobile users in $I_n^k$.

*Proof.* Assuming that $\tilde{s}$ is the NE of game $G'$ and $s^*$ is the centralized optimal solution for all the mobile users. According to the conclusion in [5] and [31], since the performance of the centralized optimal algorithm is better than that of game theory based algorithm, so $PoA \ge 1$.

For the game theory based algorithm, when $\lambda_n = 1$, the transmission rate of user $n$ satisfies that:

$$r_n(\tilde{s}) = w_k \log_2\left(1+\frac{p_n G_n}{N_0+\sum_{j\in I_n^k, \lambda_j=1} p_j G_j}\right)$$
$$\ge w_k \log_2\left(1+\frac{p_n G_n}{N_0+\sum_{j\in I_n^k, \lambda_j=1} p'_{max} G_j}\right) \quad (26)$$

where $p'_{max} \triangleq \max\{p_j, j\in I_n^k\}$. Based on (8) and (26), the overhead of user $n$ satisfies that:

$$U_{n,c} = \frac{(\alpha_t+\alpha_e p_n)s_n}{w_k \log_2\left(1+\frac{p_n G_n}{N_0+\sum_{j\in I_n^k, \lambda_j=1} p_j G_j}\right)} + \alpha_t t_{n,c} + \alpha_e e_{n,c}$$
$$\le \frac{(\alpha_t+\alpha_e p_n)s_n}{w_k \log_2\left(1+\frac{p_n G_n}{N_0+\sum_{j\in I_n^k, \lambda_j=1} p'_{max} G_j}\right)} + \alpha_t t_{n,c} + \alpha_e e_{n,c} = U_{n,c}^{max}$$
(27)

When $\lambda_n = 0$, the overhead of user $n$ can be calculated as:

$$U_n = \frac{\alpha_t c_n}{f_n^{max}} + \alpha_e \kappa_n c_n (f_n^{max})^2 \le \frac{\alpha_t c_n}{f_n^{max}} + \alpha_e \kappa_n c_n (f_n^{max})^2 = U_n^{max} \quad (28)$$

For the centralized optimal algorithm, when $\lambda_n = 1$, the transmission rate satisfies that:

$$r_n(\mathbf{s}^*) = w_k \log_2\left(1 + \frac{p_n G_n}{N_0 + \sum_{j \in I_n^k, \lambda_j=1} p_j G_j}\right) \le w_k \log_2\left(1 + \frac{p_n G_n}{N_0}\right) \quad (29)$$

Therefore, when $\lambda_n = 1$, we have:

$$U_{n,c} = \frac{(\alpha_t + \alpha_e p_n) s_n}{w_k \log_2\left(1 + \frac{p_n G_n}{N_0 + \sum_{j \in N \setminus \{n\}, \lambda_j=1} p_j G_j}\right)} + \alpha_t t_{n,c} + \alpha_e e_{n,c}$$

$$\ge \frac{(\alpha_t + \alpha_e p_n) s_n}{w_k \log_2\left(1 + \frac{p_n G_n}{N_0}\right)} + \alpha_t t_{n,c} + \alpha_e e_{n,c} = \overline{U}_{n,c}^{min} \quad (30)$$

Similar to the game theory based algorithm, when $\lambda_n = 0$, the computation overhead is:

$$U_n = \frac{\alpha_t c_n}{f_n} + \alpha_e \kappa_n c_n f_n^2 \ge \frac{\alpha_t c_n}{f_n^{min}} + \alpha_e \kappa_n c_n (f_n^{min})^2 = \overline{U}_n^{min} \quad (31)$$

Thus, according to (27), (28), (30), and (31), the Corollary 6 holds. ■

Note that the PoA reflects the worst cause of the game theory based algorithm over the centralized optimal algorithm; therefore, based on the Corollary 6, we can conclude that the PoA will decrease as the interference from the interference users reduces. This demonstrates that controlling network interference is effective in improving the performances of MEC, which is consistent with the conclusion in the Corollary 4.

**Corollary 7.** The NE of game $G'$ is Pareto Efficiency.

*Proof.* Based on [37], the Pareto efficiency is defined as: if $\mathbf{x} = \{x_1, x_2, \dots, x_N\}$, where $x_i \in \mathbb{R}$ and $i \in \mathbf{N}$, is Pareto optimal solution and $U_i$ is the utility of $i$, then there is no other feasible solution $\mathbf{x}' = \{x_1', x_2', \dots, x_N'\}$, such that $U_i(x_i') \ge U_i(x_i)$ for all the users with $U_i(x_i') > U_i(x_i)$ for some users.

In game $G'$, assuming that $\mathbf{s} = \{s_1, s_2, \dots, s_N\}$ is a NE; so for $\forall n \in \mathbf{N}$, there is no solution $\mathbf{s}' = \{s_1', s_2', \dots, s_N'\}$, which can make the $U_n(s_n, s_{-n}) \ge U_n(s_n', s_{-n})$ true; moreover, the solution $\mathbf{s}' = \{s_1', s_2', \dots, s_N'\}$ also cannot make the $U_n(s_n, s_{-n}) > U_n(s_n', s_{-n})$ true for all the users. So, on the one hand, the NE of game $G'$ is the optimal solution to the optimal function shown in (11), which means that $U_i(x_i') > U_i(x_i)$ for some users holds; on the other hand, no users can reduce their computation overhead without increasing other users overhead, which means that $U_i(x_i') \ge U_i(x_i)$ for all the users hold. And thereby, the NE of game $G'$ is Pareto efficiency. ■

## VI. NUMERICAL RESULT

In this section, we will show the performances of the proposed algorithm by simulation. In this simulation, the mobile users are randomly deployed at the coverage area of BS with the numbers vary from 8 to 20. The bandwidth of the wireless channel is 20MHz and the number of sub-channels is 10. The transmission power of user changes from $p_{min}$ to $p_{max}$; the $p_{min}$ can be calculated according to the SINR threshold and the measured interference in Section 4; the $p_{max}$ is set to 150mW. The noise is -100dBm [26]. The channel gain $G_n = d_{n,l}^{-\gamma}$ [26], where $d_{n,l}$ is the distance between the mobile user $n$ and BS $l$; $\gamma$ is the path loss factor which is set to 4. Similar to [5], in this simulation, $L_n = 5000$Kb and $C_n = 1000$Megacycles. The CPU computation capability $f_c$ is 10GHz. The decision weights $\alpha_t, \alpha_e \in [0,1]$ and $\alpha_t + \alpha_e = 1$, so we set $\alpha_t \in \{1, 0.5, 0\}$ [5]. For each mobile user, $\kappa_n = 10^{-27}$ and $f_{n,max} = 1$GHz [13].

### A. Convergence of the game

Our results show that for our proposed game theory based algorithm, the offloading decision, the transmission power, the CPU capability, and the overhead of each user can coverage to the NE point. The Fig.1(a) shows the convergence of the offloading decisions. The same as that proved in the Corollary 1, the value of λ converges to 0 or 1 with the increasing of the iteration times. As shown in the Fig.1(b), the value of CPU capability is also binary, i.e. $f = 0$ or $f = f_{max}$. Moreover, the values of the CPU capability and the offloading decision are contrary, which can be found in the Fig.1(a) and Fig.1(b). This is consistent with the theoretical analysis in Section 4. The transmission power of mobile user is shown in the Fig.1(c), which is convergent and smaller than $p_{max}$. Different from the CPU capability, when the offloading decision is 1, the transmission power is larger than 0; otherwise, the transmission power equals to 0. The Fig.1(d) demonstrates that the computation overhead of each mobile user is convergent, too.

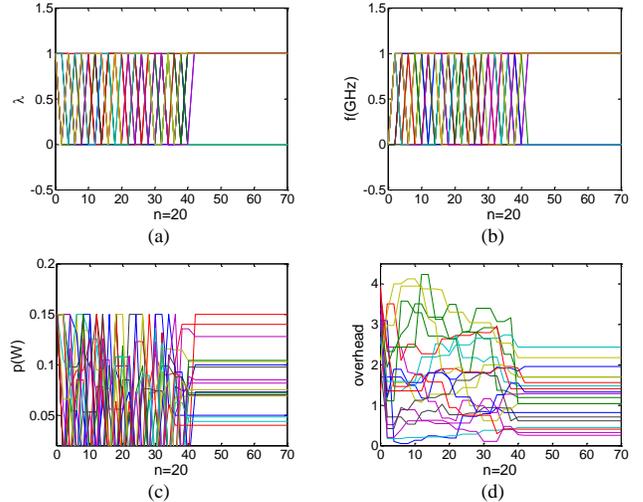

Fig.1 Convergence of the game: (a) offloading decision; (b) CPU capability; (c) transmission power; (d) computation overhead

### B. Effect of interference

Our result shows that reducing the interference can improve the performances of the proposed algorithm effectively, which is consistent with the conclusion of the Corollary 6. Fig.2 demonstrates the variation of PoA when increases the interference. In this simulation, we use $1/SINR$ to represent the effect of the interference, where $SINR = \sum_{n \in N} SINR_n$ and $SINR_n$ is the SINR of user n. So the larger $1/SINR$, the larger interference. With the increase of the interference, the PoA increases. When the value of 1/SINR is fixed, such as 0.7, the PoA reduces with the increase of the network density. This is because when the value of 1/SINR is fixed, the more users in the network, the smaller interference of each user, which also means small PoA.

### C. Effect of the number of users

Our result shows that the number of users has a great effect on the performances of the algorithm: first, the more users in the network, the more users to offload the task to the MEC server; second, the more users in the network, the higher network computation overhead is; third, the more users, the longer time to reach the NE point. The Fig.3 illustrates the number of users who offload their computation task to the MEC server under different network densities. Since the algorithm is convergent, the number of users that offload the task to MEC server becomes constant. Before reaching the NE point, with the increasing of the iteration time, the resources in mobile users are consumed and more and more mobile users choose to offload the computation task to the MEC server. This is because the MEC server has better computation capability and resources than mobile users. The number of users which offload tasks to the MEC server increases when the network density grows. However, this increase becomes slowly when the network density is large, since the interference is serious in this scenario.

network density increases, too. This is because when the task size increases, the computation overhead of the local execution increases, which can be found in (23); therefore, more users will offload their computation task to the MEC server for saving energy and reducing latency. The Fig.4 also demonstrates that when the computation load increases, the network overhead increases, too. The reason is similar to that shown in the Fig.3.

## VII. CONCLUSION

In this paper, we make the following key contributions. First, we proposed a game theory based joint offloading decision and resources allocation algorithm for multi-user MEC. Second, we prove that this game is an exact potential game and the NE of this game exists and is unique. Third, the convergence and the computational complexity of this algorithm are investigated. Fourth, we investigate the PoA of this algorithm and conclude that the interference from the interference users has the main effect on the performances of MEC, which is consistent with the conclusion in the Corollary 4. Fifth, we prove that the NE of this game is Pareto efficiency and also the global optimal solution shown in (10). The simulation results also show the effectiveness of this algorithm on improving the performances of MEC.

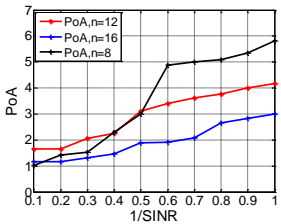
Fig.2. The PoA under different network conditions

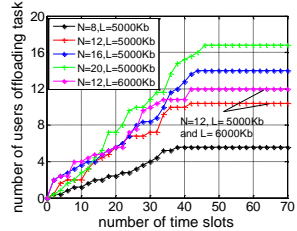
Fig.3. The number of beneficial users under different network conditions

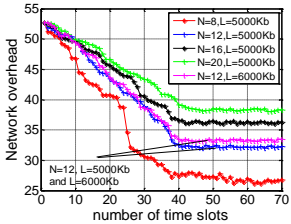
Fig.4. The network overhead under different network conditions

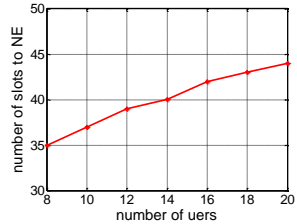
Fig.5. The convergence rate under different network density

In the Fig.4, the network overheads under different network densities are presented. Different from the results shown in the Fig.3, in the Fig.4, with the increase of the iteration time, the network overhead decreases before reaching the NE point. When the network reaches NE, the network overhead keeps stable. This result demonstrates that the proposed algorithm is effective in reducing network computation overhead. For different network densities, the more users in the network are, the higher computation overhead is. This is because the more users, the more serious interference, which leads to high network computation overhead. However, due to the game between different users, the increase of the computation overhead with the increase of the network density becomes slowly. This means that the proposed algorithm is effective in improving the performances of MEC. The number of iteration times for reaching NE under different network densities is shown in the Fig.5. Due to the interference, the more users are, the more iteration times are needed to reach the NE point. The increase is near a linear, which means that the algorithm can converge in a fast manner.

### D. Effect of task size

*Our result shows that the size of the task also has a great effect on the performances of the algorithm; the larger task size is, the more users choose to offload the task to the MEC server and the higher network computation overhead is.* As shown in the Fig. 3, if the length of input data increases, the number of users that offload the tasks to the MEC server under the same